\documentclass[showpacs,preprintnumbers,amsmath,amssymb]{revtex4}
\usepackage{bm}
\usepackage{epsfig}
\usepackage{graphicx}
\usepackage{amsmath}
\usepackage{dcolumn}
\usepackage{epstopdf}

\begin{document}

\title{Solution of the $\eta-^4$He problem with quasi-particle formalism}

\author{
A.~Fix$^{1}$\footnote{\emph{eMail address:} fix@tpu.ru} and
O.~Kolesnikov$^{2}$\thanks{\emph{eMail address:} ostrick@kph.uni-mainz.de}}
\affiliation{\mbox{$^1$Tomsk Polytechnic University, Tomsk, Russia}
\\ \mbox{$^2$Tomsk State University, Tomsk, Russia}}

\date{\today}

\begin{abstract}
The Alt-Grassberger-Sandhas equations for the five-body $\eta-4N$ problem are solved for the case of the driving $\eta N$ and $NN$ potentials limited to $s$-waves. The quasi-particle (Schmidt) method is employed to convert the equations into the effective two-body form. Numerical results are presented for the $\eta^4$He scattering length.
\end{abstract}


\maketitle

\section{Introduction}
During the last years considerable attention has been paid to interaction of $\eta$-mesons with four nucleons \cite{KruscheWilkin,Machner,Willis,Wronska, Budzanowski,Moskal,Adlarson1,Kelkar,ScRiska}. Analysis of different data is mainly focused on the search for $\eta^4$He bound states. According to the available experimental results, the rise of the $dd\to\eta^4$He experimental cross section at $E_{\eta}\to 0$ seems to be not as steep as in the $pd\to\eta^3$He reaction. As discussed, e.g., in Ref.\,\cite{Willis} the most natural interpretation of this fact is that due to additional attraction caused by one extra nucleon the pole in the $\eta^4$He scattering matrix is shifted into the region of negative values of $Re\,E_{\eta}$ and turns out to be farther from the physical region than the $\eta^3$He pole. It is therefore concluded that formation of the bound $\eta^4$He state is highly probable.

In view of general complexity of the five-body $\eta-4N$ problem
there are still no rigorous few-body calculations of this system. At the same time, a systematic practical way of handling the $n$-body interaction is provided by the quasi-particle formalism in which the kernels of integral equations are represented by series of separable terms. This method becomes especially efficient if the driving two-particle potentials are governed by the nearly lying resonances or bound (virtual) states, like in the $NN$ and $\eta N$ case. Then reasonable accuracy may be achieved with only few separable terms retained in the series. In particular, the quasi-particle formalism is shown to be very well suited for practical calculation of $\eta NN$ \cite{Shev,Pena,FiAr2N} as well as $\eta-3N$ \cite{FiAr3N} scattering (in Ref.\,\cite{Barnea} another method based on the hypospherical function expansion has been developed).

In this letter we apply the quasi-particle method to study the five-body system $\eta-4N$. As a formal basis we use the Alt-Grassberger-Sandhas $n$-body  equations derived in Ref.\,\cite{GS}. For the sake of simplicity we neglect influence of the spin and isospin on the interaction between nucleons, treating them as spinless indistinguishable particles. Furthermore, since only the threshold $\eta^4$He energies are considered, we restrict all interactions to $s$-waves only.

\section{Formalism}
As is well known, separable expansion of the kernels allows one to reduce the $n$-body integral equations to the $(n-1)$-body equations, where two of $n$ particles in each state are effectively treated as a composite particle (quasi-particle).
Therefore, the essence of the method is to approximate the $(n-1)$-particle interaction obtained in the separable-potential model again by the separable ansatz. In this respect, to simplify presentation of the formalism, we start directly from successive application of the quasi-particle technique to 2-, 3-, and 4-body subamplitudes, occurring when the five-body system is divided into groups of mutually interacting particles.

In what follows, we use the concept of partitions as introduced, e.g., in Ref.\,\cite{Yakubovsky}.
Different partitions (as well as the quasi-particles related to these partitions) are further denoted by the symbols $\alpha,\beta,\ldots$, whereas the Latin letters $a,b,\ldots$ are used for numbering the terms in the separable expansions of the subamplitudes. The notation $\alpha_n$ refers to the partition obtained by dividing the $\eta-4N$ system into $n$ groups. Writing $\alpha_{n+1}\subset \alpha_{n}$ means that the partition $\alpha_{n+1}$ is obtained from $\alpha_{n}$ via further division of the quasi-particle $\alpha_{n}$ into two groups of particles.
%============================= Fig.0 ===============================>
\begin{figure}[ht]
\begin{center}
\resizebox{0.25\textwidth}{!}{%
\includegraphics{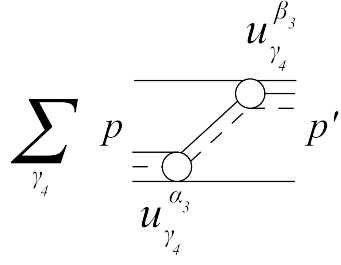}}
\caption{The potential $Z^{\alpha_2}_{\alpha_3,\beta_3}(z;p,p^\prime)$ as defined in Eq.\,(\ref{eq1_25}) connecting two configurations of the type $(\eta NN)+N$. The dashed and the solid lines represent, respectively, $\eta$-mesons and nucleons. The form factors $u^{\alpha_3}_{\gamma_4}$, $u^{\beta_3}_{\gamma_4}$ are shown by the circles.}
\label{fig0}
\end{center}
\end{figure}
%====================================================================>

The basic ingredient of the formalism is a separable expansion of the
quasi-particle amplitudes
\begin{eqnarray}\label{eq1_10}
&&X^{\alpha_{n}}_{\alpha_{n+1}a,\beta_{n+1}b}(z)=
\sum_{k,l=1}^{N_{\alpha_{n}}}|u^{\alpha_{n}(k)}_{\alpha_{n+1}(a)}
\rangle\Delta^{\alpha_{n}}_{kl}(z)\langle
u^{\alpha_{n}(l)}_{\beta_{n+1}(b)}|\,,\\
&&\alpha_{n+1},\beta_{n+1}\subset\alpha_{n}.\nonumber
\end{eqnarray}
Then the integral equations for the amplitudes $X^{\alpha_{n-1}}_{\alpha_n,\beta_n}$ are transformed exactly into the quasi-two-body equations which in the operator form read
\begin{eqnarray}\label{eq1_15}
&&X^{\alpha_{n-1}}_{\alpha_na,\beta_nb}=Z^{\alpha_{n-1}}_{\alpha_na,\beta_nb}+\sum_{\gamma_n} \sum_{k,l=1}^{N_{\gamma_n}} Z^{\alpha_{n-1}}_{\alpha_na,\gamma_nk}
\Delta^{\gamma_n}_{kl}X^{\alpha_{n-1}}_{\gamma_nl,\beta_nb}\,,\nonumber\\
&&\alpha_{n},\beta_{n},\gamma_n\subset\alpha_{n-1}\,,
\end{eqnarray}
or more explicitly
\begin{eqnarray}\label{eq1_20}
X^{\alpha_{n-1}}_{\alpha_na,\beta_nb}(z;p,p')&=&Z^{\alpha_{n-1}}_{\alpha_na,\beta_nb}(z;p,p')
+\sum_{\gamma_n\subset\alpha_{n-1}}\sum_{k,l=1}^{N_{\gamma_n}}\int\,\frac{p^{\prime\prime\,2}dp^{\prime\prime}}{2\pi^2} \, Z^{\alpha_{n-1}}_{\alpha_na,\gamma_nk}(z;p,p^{\prime\prime})\nonumber\\
&\times&\Delta^{\gamma_n}_{kl}\left(z-\frac{p^{\prime\prime\,2}}{2\mu_{\gamma_{n}}}\right)
X^{\alpha_{n-1}}_{\gamma_nl,\beta_nb}(z;p^{\prime\prime},p^\prime)
\end{eqnarray}
with $\mu_{\gamma_{n}}$ being the reduced mass associated with the partition  $\gamma_{n}$. The effective potentials $Z^{\alpha_{n-1}}_{\alpha_n,\beta_n}$ are determined as matrix elements of the 'resolvent' $\Delta^{\gamma_{n+1}}$ between the form factors appearing in the expansion (\ref{eq1_10})
\begin{eqnarray}\label{eq1_25}
&&Z^{\alpha_{n-1}}_{\alpha_na,\beta_nb}=\sum_{\gamma_{n+1}}
\sum_{k,l}\langle u^{\alpha_n(a)}_{\gamma_{n+1}(k)}|\Delta^{\gamma_{n+1}}_{kl}
|u^{\beta_n(b)}_{\gamma_{n+1}(l)}\rangle,\\
&&\gamma_{n+1}\subset\alpha_n,\ \gamma_{n+1}\subset\beta_n,\quad \alpha_n\ne\beta_n\,.\nonumber
\end{eqnarray}
The structure of Eq.\,(\ref{eq1_25}) is conveniently illustrated in the form of diagrams. In Fig.\,\ref{fig0} we show as an example one of the effective potentials $Z^{\alpha_2}_{\alpha_3,\beta_3}$, connecting two configurations of the type $(\eta NN)+N$. Since the nucleons are identical, the condition $\alpha_n\ne\beta_n$ in Eq.\,(\ref{eq1_25}) means that the nucleon lines, entering the quasi-particles $\alpha_n$ and $\beta_n$ and not included into the quasi-particle $\gamma_{n+1}$ should be different.
To calculate the form factors $u^{\alpha_n(a)}_{\gamma_{n+1}(k)}$ and the propagators $\Delta^{\gamma_{n+1}}_{kl}$ we employed the energy dependent pole expansion (EDPE) method of Ref.\,\cite{EDPE}.

\subsection{Four-body partitions}
Considering nucleons as indistinguishable particles we have
only two different types of four-particle partitions:
\begin{equation}\label{eq2_10}
  1:\  (NN)+N+N+\eta\,, \quad
  2:\  (\eta N)+N+N+N\,.
\end{equation}
The partitions 1 and 2 and the related
two-particle subsystems $NN$ and $\eta N$
will further be labeled by the index $\alpha_4=1,2$.

In the present calculation, the $NN$ and $\eta N$ $s$-wave interactions were approximated by simplest rank-one separable potentials. For $NN$ we employed
\begin{equation}\label{eq2_20}
v_1(z)=-|g_1\rangle\langle g_1|\,.
\end{equation}
The corresponding $t$-matrix has the usual form
\begin{equation}\label{eq2_21}
t_1(z)=|g_1\rangle \tau_1(z)\langle g_1|
\end{equation}
with the $NN$ propagator
\begin{equation}\label{eq2_15}
\tau_1(z)=-\bigg[1+\frac{1}{2\pi^2}\int_0^\infty\frac{g_1(q)^2}{z-q^2/M_N}\,q^2dq
\bigg]^{-1}\,,
\end{equation}
where $M_N$ is the nucleon mass. The form factors were chosen in the Yamaguchi form
\begin{equation}\label{eq2_22}
g_1(q)=\frac{\sqrt{\lambda_{NN}}}{1+(q/\beta)^2}\,.
\end{equation}
Since we treat nucleons as spinless particles, the strength $\lambda_{NN}$ was taken as an average of the singlet and the triplet strength
\begin{equation}\label{eq2_30}
\lambda_{NN}=\frac{1}{2}(\lambda^{(0)}_{NN}+\lambda^{(1)}_{NN}),\quad
\lambda^{(s)}_{NN}=\frac{8\pi a_s}{M_N(a_s\beta-2)}.
\end{equation}
The singlet and the triplet scattering lengths, $a_0$ and $a_1$, as well as the cut-off momentum $\beta$ were taken directly from the analysis \cite{Yamag} of the low-energy $np$ scattering
\begin{equation}\label{eq2_35}
a_0=23.690\,\mbox{fm},\ a_1=-5.378\,\mbox{fm},\ \beta=1.4488\,\mbox{fm}^{-1}\,.
\end{equation}
It is well known that the Yamaguchi $NN$ potential overestimates attraction at high momenta and yields significant overbinding already in the $^3$He case (see Table \ref{tab1}). Therefore we also adopted the  spin-independent $NN$ potential with exponential form factors
\begin{equation}\label{eq2_23}
g_1(q)=\sqrt{\lambda_{NN}}\,e^{-q^2/\beta^2}\,,
\end{equation}
which yields the same binding energy $E_{NN}$ of two nucleons.
The form factors (\ref{eq2_23}) with parameters listed in Table \ref{tab1} give for the three- and four-nucleon binding energies, $E_{3N}$ and $E_{4N}$, the values which are rather close to those of the $^3$He and $^4$He nuclei. At the same time, with the Gauss form factors we have a visibly larger value of the $NN$ effective range $r_0$ (see Table \ref{tab1}).

% ============================ Table 1 ==================================>
\begin{table}
\renewcommand{\arraystretch}{1.3}
\caption{The $NN$ potential parameters. $E_{NN}$, $E_{3N}$, and $E_{4N}$ are the two-, three-, and four-nucleon binding energies calculated with our model.}
\begin{tabular*}{9.1cm}%{\textwidth}
{@{\hspace{0.2cm}}c@{\hspace{0.2cm}}|@{\hspace{0.4cm}}c@{\hspace{0.5cm}}c@{\hspace{0.5cm}}
c@{\hspace{0.5cm}}c@{\hspace{0.5cm}}c@{\hspace{0.5cm}}c@{\hspace{0.2cm}}}
\hline\hline
Type & $\lambda_{NN}$  & $\beta$ & $E_{NN}$  &$r_0$ & $E_{3N}$ & $E_{4N}$ \\
  & fm$^2$ & fm$^{-1}$ & MeV & fm & MeV & MeV \\
\hline
Yamaguchi & 4.17 & 1.45 & 0.428 & 1.89 & 12.6 & 54.8 \\
Gauss & 6.51 & 1.24 & 0.428 & 2.33 & 8.05 & 30.3 \\
\hline\hline
\end{tabular*}
\label{tab1}
\end{table}
%========================================================================>

\begin{figure}[ht]
\begin{center}
\resizebox{0.5\textwidth}{!}{%
\includegraphics{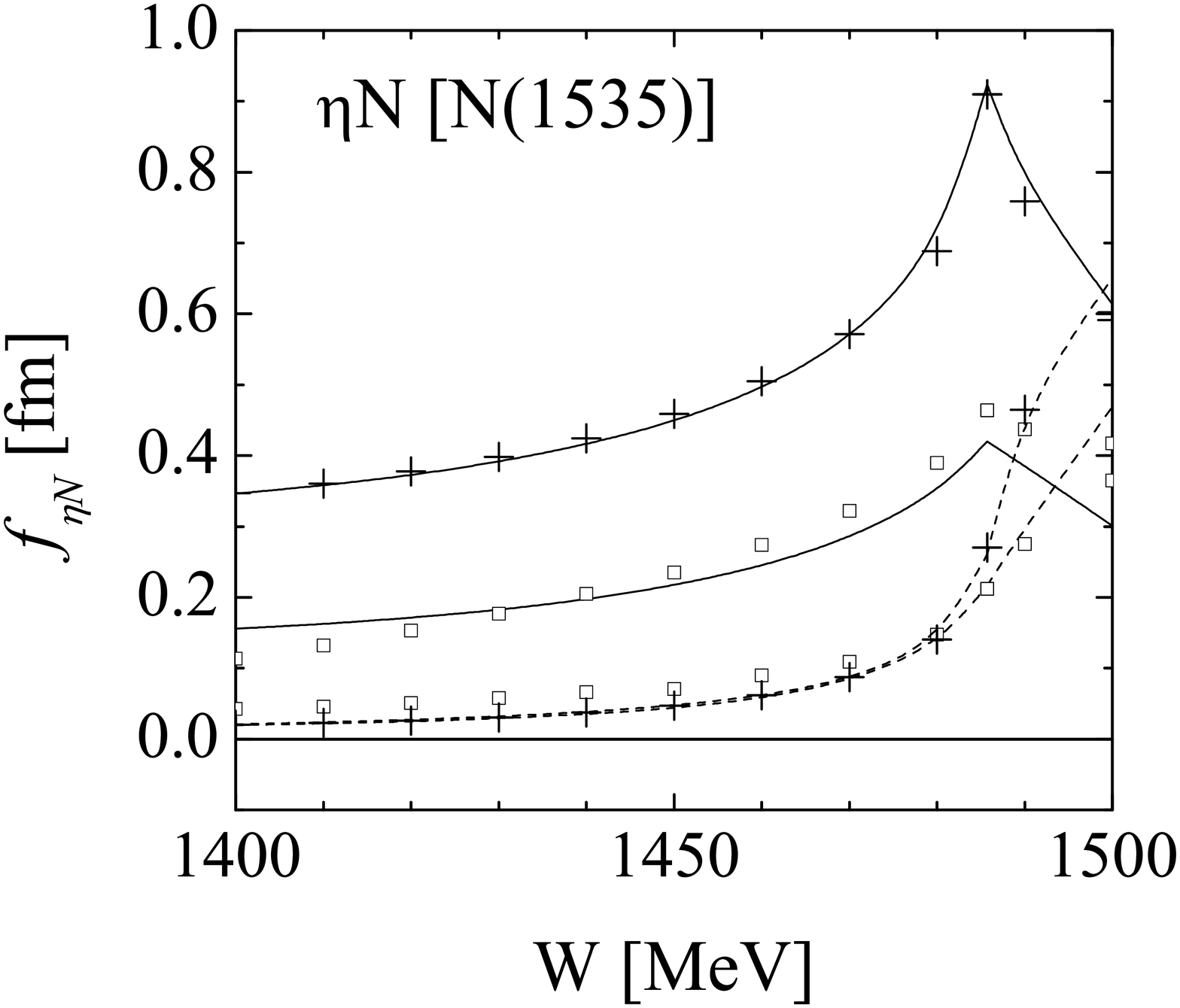}}
\caption{The $S_{11}$ partial wave of the $\eta N$ scattering
amplitude calculated with Sets I and II of the parameters listed in Table\,\ref{tab2}.
Notations: solid curve: real part,
dashed curve: imaginary part. Crosses and squares represent the results of the coupled channel analysis of Refs.\,\cite{Wycech} and \cite{KSW}, respectively.}
\label{fig1}
\end{center}
\end{figure}

The $\eta N$ $s$-wave interaction was reduced to excitation of the resonance $N(1535)1/2^-$ only. To include pions we used a conventional coupled channel formalism, where the resulting separable $t$-matrix has the matrix form
\begin{equation}\label{eq2_40}
t_{\mu\nu}(z)=\frac{1}{W-M_0}|g_\mu\rangle\tau_2(z)\langle g_\nu|\,,\quad \mu,\nu\in\{\pi,\eta\}
\end{equation}
with
\begin{equation}\label{eq2_45}
g_\mu(q)=\frac{g_\mu}{1+(q/\beta_\mu)^2}\,.
\end{equation}
The propagator
\begin{equation}\label{eq2_50}
\tau_2(z)=\frac{1}
{W-M_0-\Sigma_\eta(W)-\Sigma_\pi(W)+\frac{i}{2}\Gamma_{\pi\pi}(W)}\nonumber
\end{equation}
with $W=z+M_N+M_\eta$, where $M_\eta$ is the $\eta$ mass, is determined by the $N(1535)1/2^-$ self-energies
$\Sigma_\eta(W)$ and $\Sigma_\pi(W)$.
The two-pion channel was included via the $\pi\pi N$ decay width $\Gamma_{\pi\pi}$  parametrized in the form
\begin{equation}\label{eq2_55}
\Gamma_{\pi\pi}(W)=\gamma_{\pi\pi}\frac{W-M_N-2M_\pi}{M_\pi}\,.
\end{equation}
The parameters $g_{\eta}$, $\beta_{\eta}$, $g_{\pi}$, $\beta_{\pi}$, $M_0$, and $\gamma_{\pi\pi}$
were chosen in such a way that the scattering amplitude $f_{\eta N}$ corresponding to our $t$-matrix $t_{\eta\eta}$ (\ref{eq2_40}) is close to that obtained in the coupled-channel analyses in the energy region from 20 MeV above the $\eta N$ threshold to 100 MeV below the threshold. Here we took the results of two works \cite{Wycech} and \cite{KSW} predicting rather different values of $Re\,f_{\eta N}$ (see Fig.\,\ref{fig1}).

% ============================ Table 2 ==================================>
\begin{table}
\renewcommand{\arraystretch}{1.3}
\caption{The $\eta N-\pi N$ parameters.}
\begin{tabular*}{9.1cm}%{\textwidth}
{@{\hspace{0.3cm}}c@{\hspace{0.3cm}}|@{\hspace{0.4cm}}c@{\hspace{0.5cm}}c@{\hspace{0.5cm}}
c@{\hspace{0.5cm}}c@{\hspace{0.5cm}}c@{\hspace{0.5cm}}c@{\hspace{0.2cm}}}
\hline\hline
Set [Ref.] & $g_{\eta}$  & $\beta_\eta$ & $g_{\pi}$  &$\beta_\pi$ & $M_0$ & $\gamma_{\pi\pi}$ \\
  &  & MeV & & MeV & MeV & MeV \\
\hline
I  \cite{Wycech} & 1.91 & 636 & 0.651 & 850 & 1577 & 4.0 \\
II  \cite{KSW} & 1.23 & 636 & 1.28 & 350 & 1527 & 1.0 \\
\hline\hline
\end{tabular*}
\label{tab2}
\end{table}
%========================================================================>

%============================= Fig. 1 ===============================>
%=====================================================================>

\subsection{Three-body partitions}
We have four different three-body partitions
\begin{equation}\label{eq3_10}
\begin{array}{llll}
  1: & (NNN)+N+\eta\,,      & 2: &  (\eta NN)+N+N\,, \\
  3: & (\eta N)+(NN)+N\,,\phantom{xxx}   & 4: &  (NN)+(NN)+\eta
\end{array}
\end{equation}
which in the following are numerated by the index $\alpha_3=1,\ldots,4$.
In the latter two cases there are two pairs of interacting particles propagating independently.
The effective potentials $Z^{\alpha_3}_{\alpha_4,\beta_4}$ determined by  Eq.\,(\ref{eq1_25}) for $n=4$ are matrix elements of the free resolvent $G_0$ between the form factors $g_{\alpha_4}$ $(\alpha_4=1,2)$
\begin{equation}\label{eq3_15}
Z^{\alpha_3}_{\alpha_4,\beta_4}=\langle g_{\alpha_4}|G_0|g_{\beta_4}\rangle\,.
\end{equation}
The functions $g_{\alpha_4}(q)$ are given by Eqs.\,(\ref{eq2_22}) (or (\ref{eq2_23})) and (\ref{eq2_45}) with $g_2(q)\equiv g_\eta(q)$. Here we omit the superfluous indices $a,b$, since our separable ansatz for $NN$
and $\eta N$ amplitudes contains in both cases only one term (see Eqs.\,(\ref{eq2_21}) and (\ref{eq2_40})).

%%============================= Fig. 2 ===============================>
\begin{figure}[ht]
\begin{center}
\resizebox{0.5\textwidth}{!}{%
\includegraphics{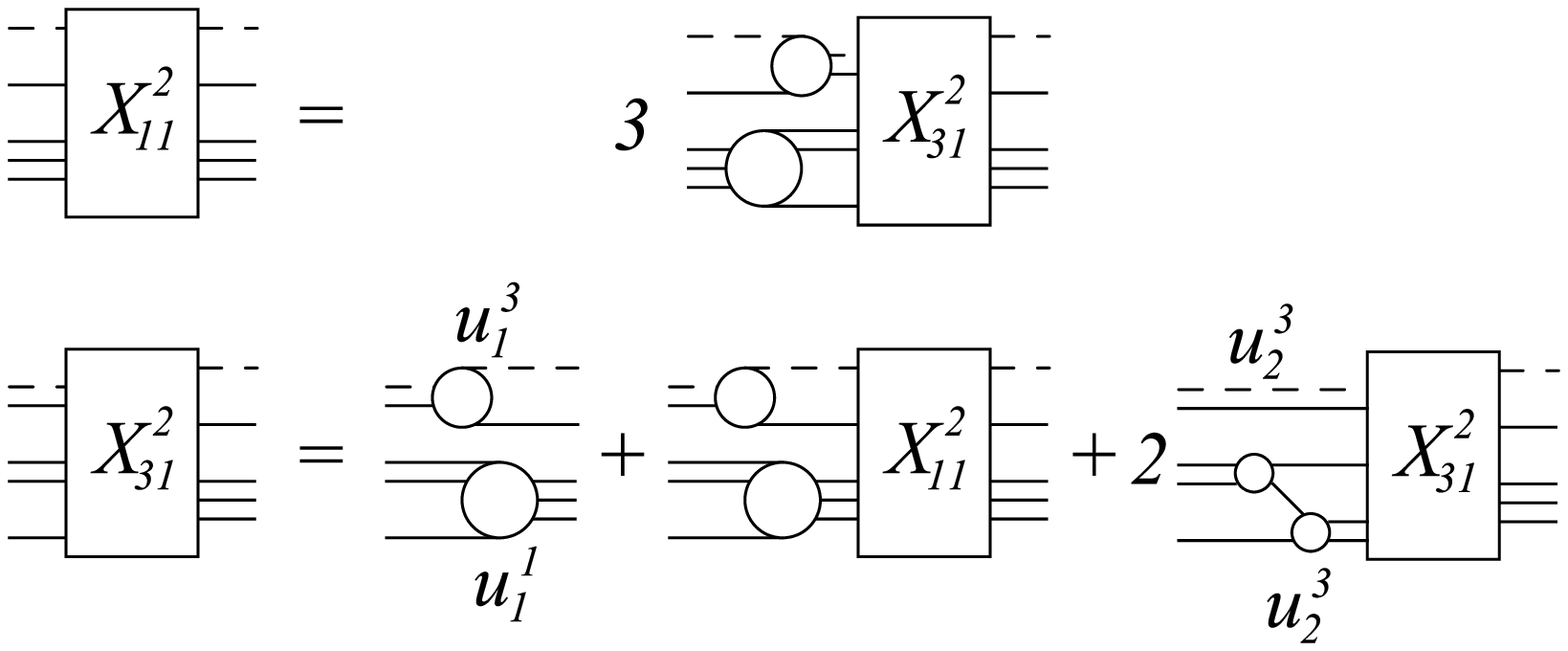}}
\caption{Effective quasi-two-body equations for the $(\eta N)-(NNN)$ amplitudes $X^2_{\alpha_3,\beta_3}$. Notation of the lines as in Fig.\,\ref{fig0}. The lower and the upper indices in $u^{\alpha_3}_{\alpha_4}$ refer to the numbers of the four- and three-body partitions listed in Eqs.\,(\ref{eq2_10}) and (\ref{eq3_10}), respectively.
The numerical coefficients appear due to symmetrization of the nucleon states.
}
\label{fig2}
\end{center}
\end{figure}
%=====================================================================>
%%============================= Fig. 3 ===============================>
\begin{figure}[ht]
\begin{center}
\resizebox{0.5\textwidth}{!}{%
\includegraphics{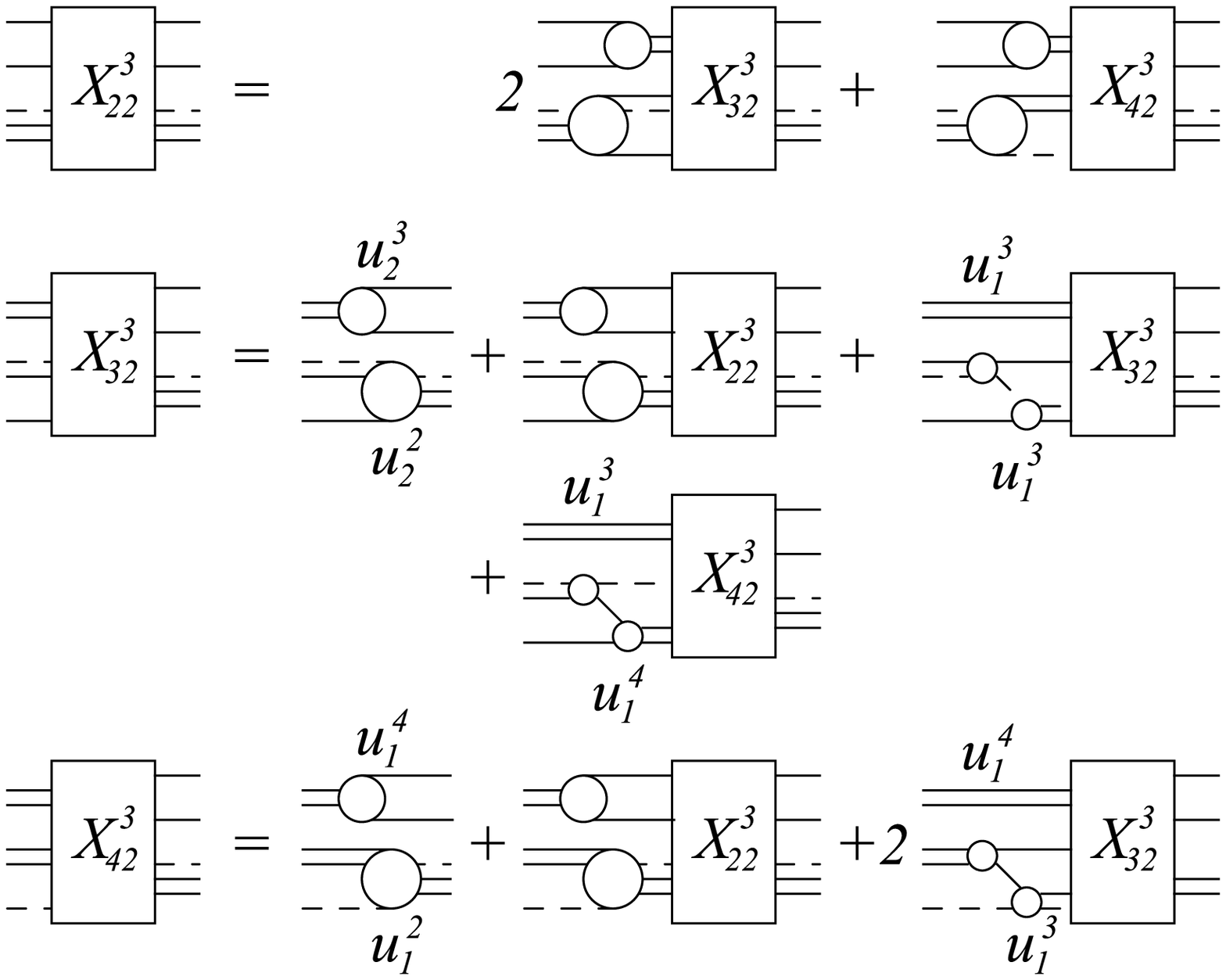}}
\caption{Same as in Fig.\,\ref{fig2} for the $(\eta NN)-(NN)$ amplitudes $X^3_{\alpha_3,\beta_3}$.}
\label{fig3}
\end{center}
\end{figure}
%=====================================================================>

\subsection{Two-body partitions}
There are four two-body partitions of the $\eta-4N$ system:
\begin{equation}\label{eq4_10}
\begin{array}{llll}
  1: & \eta+(NNNN)\,,      & 2: & (\eta N)+(NNN)\,,  \\
  3: & (\eta NN)+(NN)\,,\phantom{xxxxx}   & 4: &  (\eta NNN)+N
\end{array}
\end{equation}
which will be labeled by $\alpha_2=1,\ldots,4$.

The effective potentials $Z^{\alpha_2}_{\alpha_3a,\beta_3b}$ are matrix elements of the 'resolvent' $\tau_{\alpha_4}$ between the form factors $u^{\alpha_3(a)}_{\alpha_4}$ appearing in the separable expansion (\ref{eq1_10}) for $n=3$:
\begin{equation}\label{eq4_15}
Z^{\alpha_2}_{\alpha_3a,\beta_3b}=\sum_{\gamma_4=1,2}
\langle u^{\alpha_3(a)}_{\gamma_4}|\tau_{\gamma_4}|u^{\beta_3(b)}_{\gamma_4}\rangle\,.
\end{equation}
The propagators $\tau_{\alpha_4}$ ($\alpha_4=1,2$) are given by (\ref{eq2_15}) and (\ref{eq2_50}) with $\tau_2\equiv\tau_{\eta\eta}$.

The calculation of the $NNNN$ ($\alpha_2=1$) and $\eta NNN$ ($\alpha_2=4$) amplitudes with separable $NN$ potentials may be found, e.g., in Refs.\,\cite{AGS_4N} and \cite{FiAr3N}, and we refer the reader to these works. The effective $(3+2)$ amplitudes ($\alpha_2=2,3$) describe propagation of two groups of mutually interacting particles. The corresponding integral equations are schematically presented in Figs.\,\ref{fig2} and \ref{fig3}.

\begin{figure*}[ht]
\begin{center}
\resizebox{1.0\textwidth}{!}{%
\includegraphics{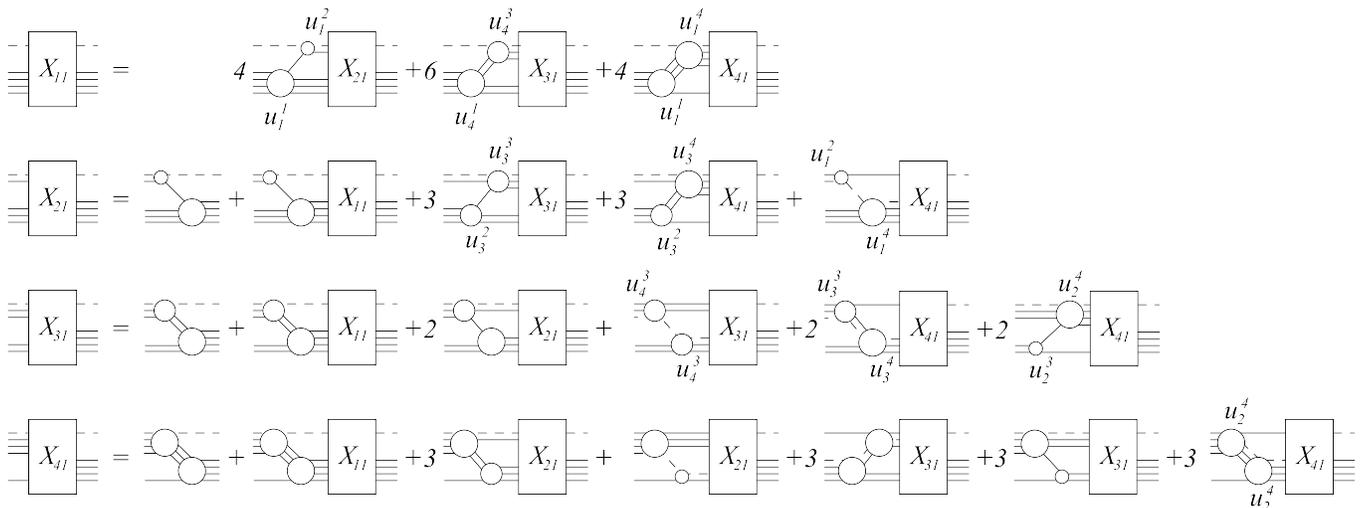}}
\caption{Graphical representation of the effective quasi-two-body equations for $\eta-4N$ scattering. Notations as in Fig.\,\ref{fig0}.
The lower and the upper indices in $u^{\alpha_2}_{\alpha_3}$ refer to the three- and two-body partitions, as given  in Eqs.\,(\ref{eq3_10}) and (\ref{eq4_10}). The numerical factors arise from the identity of the nucleons.}
\label{fig4}
\end{center}
\end{figure*}

After the separable expansions (\ref{eq1_10}) for $n=2$ are calculated
we build the effective potentials $Z_{\alpha_2a,\beta_2b}$ (\ref{eq1_25}) as
\begin{equation}\label{eq4_30}
Z_{\alpha_2a,\beta_2b}=\sum_{\gamma_3=1}^4\sum_{k,l=1}^{N_{\gamma_3}}
%\Gamma_{\alpha_2\beta_2;\gamma_3}
\langle
u^{\alpha_2(a)}_{\gamma_3(k)}|\Delta^{\gamma_3}_{kl}
|u^{\beta_2(b)}_{\gamma_3(l)}\rangle\,.
\end{equation}
The corresponding system of the five-body $\eta-4N$ equations is diagrammatically presented in Fig.\,\ref{fig4}.
After this system is solved, the $\eta ^4$He scattering amplitude can be calculated as
\begin{equation}\label{eq4_45}
f_{\eta^4\mathrm{He}}(p)=-N^2\frac{\mu}{2\pi}\,X_{11,11}(z;p,p)\,.
\end{equation}
Here $N$ is the normalization constant of the $^4$He wave function, $\mu$ is the $\eta-^4$He reduced mass, and the momentum $p$ is fixed by the on-mass-shell condition
\begin{equation}\label{eq4_50}
p=\sqrt{2\mu (z+E_{4N})}\,,
\end{equation}
where $E_{4N}>0$ is the four-nucleon binding energy given in Table \ref{tab1}.
%============================= Fig. 4 ===============================>
%=====================================================================>

% ============================ Table 3 ==================================>
\begin{table}
\renewcommand{\arraystretch}{1.5}
\caption{
The scattering length $a_{\eta^4\mathrm{He}}$ as a function of
$N_{\alpha_2}$ ($\alpha_2=1,\ldots,4$), the number of separable terms retained in the separable expansion
(\ref{eq1_10}) for the (4+1) and (3+2) subamplitudes $X^{\alpha_2}$. The calculation is performed with the Gauss $NN$ potential and Set I of the $\eta N-\pi N$ parameters.}
\begin{tabular*}{8.3cm}%{\textwidth}
{@{\hspace{0.6cm}}c@{\hspace{0.9cm}}c@{\hspace{0.9cm}}
c@{\hspace{0.9cm}}c@{\hspace{0.9cm}}
|@{\hspace{0.6cm}}c@{\hspace{0.9cm}}}
%\begin{tabular}{ccccc|c}
\hline\hline
$N_1$ & $N_2$ & $N_3$ & $N_4$ & $a_{\eta\, ^4\mathrm{He}}$ [fm] \\ \hline
 2    &  2    &   2    &   2    &   $5.56+0.96\,i$ \\
 4   &   4    &   4    &   4    &   $4.88+1.23\,i$ \\
 4   &   4    &   6    &   6    &   $4.83+1.23\,i$ \\
 6   &   6    &   8    &   8    &   $4.79+1.22\,i$ \\
% 8   &   8    &   10   &   10   &   $4.79+1.22\,i$ \\
10   &   10   &   12   &   12   &   $4.79+1.22\,i$ \\
% 15   &   15   &   15   &   15   &   $4.80+1.22\,i$ \\
20   &   20   &   20   &   20   &   $4.80+1.22\,i$ \\
\hline\hline
\end{tabular*}
\label{tab3}
\end{table}
%========================================================================>
In Table\,\ref{tab3} we present the value of the $\eta^4$He scattering length
calculated with different number $N_{\alpha_2}$ of terms retained in the separable expansion (\ref{eq1_10}) of the amplitudes $X^{\alpha_2}_{\alpha_3,\beta_3}$. As one can see, satisfactory accuracy is achieved with $N_1=N_2=6$, $N_3=N_4=8$. In principle, already with first four terms in each expansion the resulting scattering length is within less than 2$\%$ of the correct value. Thus, also in the five-body case $\eta-4N$ the quasi-particle approach based on the EDPE method of Ref.\,\cite{EDPE} is very suitable for practical applications. The minimum number of separable terms $N_{\alpha_2}$ only slightly exceeds that for the four-body kernels, where convergence is achieved already with first four-six terms in each subamplitude.

\section{Discussion and conclusion}
\label{sec:results}

As our main result we present the $\eta^4$He scattering length $a_{\eta^4\mathrm{He}}=f_{\eta^4\mathrm{He}}(0)$. It is given in Table \ref{tab4} for two versions of the $NN$ potential. For comparison purposes also the $\eta^3$He scattering length calculated with the same sets of the $NN$ and $\eta N-\pi N$ parameters is presented.

It is remarkable, that despite the larger number of nucleons in $^4$He
the predicted value of $a_{\eta^4\mathrm{He}}$ is smaller than $a_{\eta^3\mathrm{He}}$. Direct calculation shows that the main reason of this somewhat unexpected result is rather rapid decrease of the $\eta N$ scattering amplitude in the subthreshold region (see Fig.\,\ref{fig1}). Because of essentially stronger binding of $^4$He in comparison to $^3$He, in the former case the effective in-medium $\eta N$ interaction acts at lower internal $\eta N$ energies, thus leading to general reduction of the attractive $\eta N$ forces (this question was addressed in detail in Refs.\,\cite{WyGrNisk,HaidLiu,WycechKrz}). This effective weakening may qualitatively explain why the peculiar slope in the $\eta$ spectrum at low energies seen in the data for $dd\to\eta^4$He \cite{Adlarson1} and $pd\to\eta ^3$He \cite{Mersmann,Smyrski} becomes less steep, when we turn from $\eta^3$He to $\eta^4$He.

% ============================ Table 4 ==================================>
\begin{table}[h]
\renewcommand{\arraystretch}{1.5}
\caption{The $\eta^3$He and $\eta^4$He scattering lengths predicted by our calculation. The first and the second rows for each version of the $NN$ potential list the values obtained with Set I and Set II of the $\eta N-\pi N$ parameters, respectively.}
\begin{tabular*}{8.3cm}
{@{\hspace{0.2cm}}c@{\hspace{0.2cm}}c@{\hspace{0.2cm}}|@{\hspace{0.8cm}}c@{\hspace{1.0cm}}
c@{\hspace{0.5cm}}}
\hline\hline
$NN$ & $\eta N-\pi N$ & $a_{\eta^3\mathrm{He}}$ [fm] & $a_{\eta^4\mathrm{He}}$ [fm] \\
\hline
Yamaguchi & I & $6.5+3.6\,i$ & $2.2+0.3\,i$ \\
      & II & $1.1+0.5\,i$ & $0.5+0.1\,i$ \\
Gauss & I & $6.7+4.0\,i$ & $4.8+1.2\,i$ \\
      & II & $1.3+0.7\,i$ & $1.0+0.3\,i$ \\
\hline\hline
\end{tabular*}
\label{tab4}
\end{table}
%========================================================================>

Summarizing, $\eta^4$He interaction is calculated for the first time
correctly dealing with the few-body aspects of the problem. Applying separable representation firstly to the $(3+1)$ and $(2+2)$ and then to the $(4+1)$ and $(3+2)$ kernels we have solved the five-body Alt-Grassberger-Sandhas equations reducing them to a coupled set of quasi-two-body equations having Lippmann-Schwinger structure.

The predicted value of $Re\,a_{\eta^4\mathrm{He}}$ is positive and turns out to be smaller than $Re\,a_{\eta^3\mathrm{He}}$. This finding should be attributed to effective weakening of the in-medium $\eta N$ interaction. According to our calculation, increase of the attractive forces due to an extra nucleon in $^4$He is overwhelmed by stronger suppression of the subthreshold $\eta N$ interaction in a more dense nucleus. The resulting attraction in the $\eta-4N$ system is too weak and does not support existence of the $\eta^4$He bound state, at least with the $\eta N$ parameters, used in the present calculation. This might be the key reason why no signal of $\eta^4$He bound state formation is still revealed, e.g., in the $dd\to^3$He\,$n\pi^0$ and $dd\to^3$He\,$p\pi^-$ reactions \cite{Krzemien,Adlarson2}.

Finally, we note that although our results obviously suffer from oversimplified treatment of the $NN$ potential, they demonstrate applicability of the quasi-particle formalism to the five-body $\eta^4$He problem. The EDPE method provides rather rapid convergence of the separable expansion, so that transition from $\eta-3N$ to the $\eta-4N$ case is performed without drastic increase of numerical complexity.
At the same time, more refined treatment requires inclusion of the nucleon spin as well as more sophisticated nucleon-nucleon potential instead of our simple rank-one ansatz.

\end{document}